\documentclass[a4paper,11pt]{article}

\usepackage{jheppub} 

\usepackage{fancyhdr}
\usepackage{graphicx}
\usepackage{xspace}
\usepackage{rotating}
\usepackage{braket}

\DeclareSymbolFont{extraup}{U}{zavm}{m}{n}
\DeclareMathSymbol{\varheart}{\mathalpha}{extraup}{86}
\DeclareMathSymbol{\vardiamond}{\mathalpha}{extraup}{87}

\usepackage{amsfonts}
\usepackage{pifont}

\newcommand{\be}{\begin{equation}}
\newcommand{\ee}{\end{equation}}
\newcommand{\bea}{\begin{eqnarray}}

\newcommand{\eea}{\end{eqnarray}}

\newcommand{\Rmnum}[1]{\expandafter\@slowromancap\romannumeral #1@}

\def\mev{\,{\rm MeV}}
\def\gev{\,{\rm GeV}}

\begin{document}

 \title{\boldmath %
 Gravitational wave and collider implications of electroweak baryogenesis aided by non-standard cosmology}
\author[1]{Micha{\l} Artymowski}
\author[2, 3]{Marek  Lewicki}
\author[3, 4]{James D. Wells}
\affiliation[1]{
Jagiellonian University,
{\L}ojasiewicza 11, 30-348 Krak{\'o}w, Poland}
\affiliation[2]{
Faculty of Physics, University of Warsaw ul.\ Pasteura 5, 02-093 Warsaw, Poland}
\affiliation[3]{
Michigan Center for Theoretical Physics, University of Michigan, Ann Arbor MI 48109, USA} 
\affiliation[4]{
Deutsches Elektronen-Synchrotron DESY, Theory Group, D-22603 Hamburg, Germany
}
\emailAdd{Michal.Artymowski@uj.edu.pl}
\emailAdd{marek.lewicki@fuw.edu.pl}
\emailAdd{jwells@umich.edu}
\abstract{
We consider various models realizing baryogenesis during the electroweak phase transition (EWBG). Our focus is their possible detection in future collider experiments and possible observation of gravitational waves emitted during the phase transition.
We also discuss the possibility of a non-standard cosmological history which can facilitate EWBG. We show how acceptable parameter space can be extended due to such a modification and conclude that next generation precision experiments such as the ILC will be able to confirm or falsify many models realizing EWBG.
We also show that, in general, collider searches are a more powerful probe than gravitational wave searches. However, observation of a deviation from the SM without any hints of gravitational waves can point to models with modified cosmological history that generically enable EWBG with weaker phase transition and thus, smaller GW signals. } 


\maketitle
\flushbottom


\section{Introduction} \label{sec:intro}

Discovery of the Higgs boson, with a mass of $125$ GeV at the Large Hadron Collider (LHC)
\cite{Aad:2012tfa,Chatrchyan:2012xdj} finally confirmed that the electroweak symmetry is broken due to a vacuum expectation value of an elementary scalar. This discovery also marks the beginning of a new era of precision measurements of the Higgs properties as a probe for physics beyond the standard model. 
Another very important recent discovery of the first gravitational wave signal \cite{Abbott:2016blz} opened a new way of probing violent events in the history of our universe through observation of the gravitational waves they would leave behind.
With these experimental prospects it is very interesting to re-examine paradigms that predict observable effects in both these areas.

In this paper, we wish to study electroweak baryogenesis \cite{Kuzmin:1985mm,Cohen:1993nk,Riotto:1999yt,Morrissey:2012db} in which a strong first order electroweak phase transition (EWPT) is responsible for the observed baryon asymmetry of the universe.
In the Standard Model (SM) the phase transition is second order with the observed Higgs mass \cite{Arnold:1992rz,Kajantie:1996qd} and so a modification is required.
We will study a simple toy model where a single new scalar is added to the SM, and we will consider several possible charge assignments for this new particle \cite{Curtin:2014jma,Cohen:2012zza,Katz:2014bha}. 
Such a modification creates a barrier between the symmetric minimum and the new electroweak symmetry breaking minimum which develops as the temperature of the universe drops, making the phase transition more strongly first order. 
This has two effects. First, modification of the high temperature potential inevitably leads to  a modification of the zero temperature Higgs potential which we can probed in colliders. And second, a more violent phase transition (i.e., stronger first order) results in larger production of gravitational waves

The main point we wish to make comes from the fact that early cosmological evolution of the universe is rather poorly constrained by experiments. To be more specific, in our discussion we will include the possibility that the early universe energy density was dominated by a new contribution not interacting with the SM which red-shifted away before nucleosynthesis. This scenario is much different from the standard assumption that the universe was dominated by radiation; however, as we will show, no currently available experimental data can exclude this possibility.

The necessary condition for baryogenesis we will address comes from the fact that the same sphaleron processes that can be responsible for creation of the asymmetry can also wash it away when the universe goes back to thermal equilibrium and the sphalerons are not sufficiently decoupled.
As mentioned already we will discuss not only how generating a larger potential barrier helps in damping the sphaleron processes but also discuss how their cosmological freeze-out can help ameliorate the situation \cite{Joyce:1996cp,Joyce:1997fc,Lewicki:2016efe,Servant:2001jh}.

While we will not discuss generation of the baryon asymmetry during the phase transition,
additional problems can appear when considering the CP violation that is also needed for the asymmetry. 
Helpful sources of CP violation are limited by increasingly accurate experimental EDM constraints~\cite{Hudson:2011zz,Harris:1999jx}, which in turn requires a stronger first-order phase transition for the asymmetry to develop~\cite{Konstandin:2013caa}.
This problem, however, is very model dependent and in some models can be completely decoupled from the sphaleron bound. Thus we will only discuss the latter as a more robust requirement.

Gravitational waves were widely discussed as a possible probe of electroweak baryogenesis \cite{Grojean:2006bp} including their interplay with collider signals \cite{Huang:2016cjm,Kakizaki:2015wua,Hashino:2016rvx,Hashino:2016xoj,Kobakhidze:2016mch,Chala:2016ykx} and possible non standard cosmological events during the phase transition \cite{Chung:2010cb,Addazi:2016fbj}. We reinvestigate these signals in our model. Strength of the GW signal drops quickly as the transition becomes weaker and generically modification of precision Higgs observables probes a larger part of the parameter space. In regions where baryogenesis is allowed due to our cosmological modification the GW signal is too weak for observation in planned searches even before considering the diminishing of the signal due to the modification.    

The simplest possible origin of our additional energy component is an oscillating homogeneous scalar field with non-renormalizable potential, i.e.\ with $V(\phi) \propto \phi^{2n}$. In that case the energy density of $\phi$ would redshift as $a^{-6n/(1+n)}$, which in the $n \gg 1$ limit gives $a^{-6}$. Such a field could originally play the role of one of the inflatons, which is very weakly coupled to the SM particles and therefore has not contributed significantly to the process of reheating. 
It is not to be confused with the new scalar that modifies the SM Higgs potential to produce a first order phase transition.
 Note, that non-renormalizable potentials are perfectly consistent with the CMB data assuming that the inflaton was non-minimally coupled to gravity \cite{Kallosh:2013tua}.

\section{Modifying The Standard Model}\label{sec:part_model}

In this section we describe our model on the particle physics side. Our starting point is
the Standard Model with the standard potential 
\begin{equation}\label{eqn:SMpot}
V(H)=-\mu^2|H|^2+\lambda|H|^4,
\end{equation}
with
$H^T=\left(\chi_1+i\chi_2,\varphi+i\chi_3\right)/\sqrt{2}$.
We assume only the real part of the neutral component has a vev: $\varphi=\phi+v$ and $\phi$ is the physical Higgs boson. 
Our modification is simply addition of a new scalar field $S$ with the potential 
\begin{equation}\label{eqn:Modpot}
V(H,S)=m_0^2|S|^2+g_S |H|^2 |S|^2+\lambda_S |S|^4,
\end{equation}
The field-dependent masses are identical to the Standard Model, and the new scalar mass takes the form
\begin{equation}
m^2_S(\phi) =m_0 + \frac{g_S}{2}\phi^2
\label{eqn:massh}
\end{equation}
 and we denote the physical mass $m_{S}=m_S(\phi=v_0)$.
We will consider several different scenarios for the charge assignment of the new scalar $S$.
 In the first simplest case it will be a singlet and can be thought of as a toy model for Higgs portal phenomenology.
In the second case it will be a color triplet and an $SU(2)$ singlet reminiscent of a right handed stop squark in the MSSM.
Details and various constants we will need in further calculations are summarized in Table~\ref{tab:NewScalar}.

\begin{table}[t]
\centering
\begin{tabular}{l | c | c | c | c | c | c | c | c}
Scenario & $SU(3)$ & $SU(2)$ & $U(1)$ & $n_S$ & $\bar{n}_S$ & $\Delta \Pi_W$ & $\Delta \Pi_B$  \\
\hline
1. Singlet & $1$ & $1$ & $1$ & $1$ & $1$ & $\frac{11}{6}$ & $\frac{11}{6}$  \\
2. RH stop & $\bar{3}$ & $1$ & $-\frac{2}{3}$ & $3$ & $3$ & $\frac{11}{6}$ & $\frac{107}{54}$ 
\end{tabular}
\caption{Charge assignments and various constants for our scenarios of the new scalar.}
\label{tab:NewScalar}
\end{table}

Following the prescription from \cite{Delaunay:2007wb},
we include thermal and loop corrections as follows,
\begin{equation}\label{eqn:Veff}
\begin{split}
V_{eff}(\phi,T)&=-\frac{m^2}{2}\phi^2+\frac{\lambda}{4}\phi^4
          + \sum_{i=h,\chi,W,Z,t,S}n_i\frac{m_{i}^4(\phi)}{64\pi^2}\left[\log\frac{m^2_{i}(\phi)}{\mu^2}-C_i\right] 
            \\
 &  +\sum_{i= h,\chi,W,Z,S}\frac{n_iT^4}{2\pi^2} J_b\left(\frac{m^2_i(\phi)}{T^2}\right)+\sum_{i= t}\frac{n_iT^4}{2\pi^2} J_f\left(\frac{m^2_i(\phi)}{T^2}\right) 
  \\
 &  +\sum_{i=h,\chi,W,Z,\gamma,S}\frac{\bar{n}_iT}{12\pi}\left[m^3_i(\phi)-\left(m^2_i(\phi)+\Pi_i(T)\right)^{3/2}\right].
\end{split}
\end{equation}
The coefficients read $n_{\{h,\chi,W,Z,t\}}=\{1,3,6,3,-12\}$, $\bar{n}_{\{h,\chi,W,Z,t\}}=\{1,3,2,1,1\}$, $C_i=3/2$ for $i=h,\chi,t,S$ and $C_i=5/6$ for $i=W,Z$, while coefficients for the new scalar are listed in Table~\ref{tab:NewScalar}.
The functions $J$ are given by 
\begin{equation}
J_{b/f }\left(\frac{m^2_i(\phi)}{T^2}\right)=\int_0^\infty dk \, 
k^2\log\left[1\mp {\rm exp}\left(-\sqrt{\frac{k^2+m_i^2(\phi)}{T^2}} \right) \right].
\end{equation}
Resumming the so called {\it daisy} diagrams we obtain the thermally corrected masses \cite{Katz:2014bha,Comelli:1996vm,Carrington:1991hz}
\begin{equation}
\begin{split}
\Pi_S (T) & =T^2 \left( 
\frac{4}{3} \frac{g_3^2}{4} +\left(-\frac{2}{3}\right)^2 \frac{g'^2}{4} + \frac{g_S}{6} + \frac{\lambda_S}{6} (2 n_S + 1)
 \right)  \nonumber \\
\Pi_{h}(T)& =T^2 \left(\frac{1}{16}g'^2 + \frac{3}{16} g^2 + \frac{\lambda}{2} + \frac{1}{4}y_t^2 + n_S \frac{g_S}{6} \right) \\
\Pi_{\chi_i}(T)& =T^2 \left(\frac{1}{16}g'^2 + \frac{3}{16} g^2 + \frac{\lambda}{2} + \frac{1}{4}y_t^2 \right) \\
\Pi_W(T)&=\Delta \Pi_W g^2 T^2\\ 
\end{split}
\end{equation}
and the shifted masses of $Z$ and $\gamma$ ($m^2_{Z/\gamma}+\Pi_{Z/\gamma}(T)$) are eigenvalues of the following mass matrix, including thermal corrections
\begin{equation}
\begin{pmatrix}
\frac{1}{4} g^2 \phi^2+ \Delta\Pi_W g^2T^2 & -\frac{1}{4}g'^2 g^2 \phi^2 \\
-\frac{1}{4}g'^2 g^2 \phi^2 & \frac{1}{4} g'^2 \phi^2+\Delta\Pi_B g'^2T^2
\end{pmatrix},
\end{equation}
with the parameters $\Delta \Pi$ listed in Table \ref{tab:NewScalar} for both discussed models. 
The values of the SM parameters $\lambda$ and $m$ are calculated from constraints on the Higgs potential 
\begin{equation}\label{masseq}
\begin{split} 
V'_{eff}(\phi,T=0)|_{\phi=v_0}=0, \quad V''_{eff}(\phi,T=0)|_{\phi=v_0}=m_h,
\end{split}
\end{equation}
which corresponds to requiring the correct prediction of the observed masses of the Higgs boson $m_h=125\gev$ and the gauge bosons via the Higgs ground state of 
$v_0 := \langle \phi (T=0) \rangle = 246\gev$.



\section{Higgs precision measurements}
In both of our models direct detection can prove difficult. In the neutral scalar case the new particle is not produced in proton collision and even future $pp$ colliders would not give stringent constraints \cite{Curtin:2014jma}. The colored case requires a bit more consideration since at first glance it should be very easily produced and detected in a pp collider. Hovewer, considering the possible decay channels one can always obscure such modification in a detector for example in a ``diquark" setup where they would always be produced in pairs \cite{Giudice:2011ak} or in the ``stealth stop" region if it is a true stop of the MSSM \cite{Aad:2012xqa,Chatrchyan:2013xna}. While these more contrived scenarios require some additional structure, we can still safely conclude that direct detection of new states is not a robust probe of EWBG scenarios.

This is not the case in Higgs precision measurements since any attempt at obfuscation of the signal has to bring our potential closer to the SM one and further from realizing EWBG.  
Also here the singlet scalar case proves to be somewhat problematic since the only measured Higgs property, modified in this model, is the triple-Higgs coupling given by the third derivative of the zero temperature potential \eqref{eqn:Veff},
\begin{equation}
\lambda_3 = \frac{1}{6} \left. \frac{d^3 V_{eff}(h,{T=0})}{d h^3} \right|_{h=v_0}.
\end{equation}
This coupling can only be measured at colliders in double Higgs production events. However very low cross-section for such events makes the measurement difficult.
High luminosity LHC is estimated to be able to determine the value of $\lambda_3$ with about $30\%$ accuracy \cite{Goertz:2013kp}.  
Future experiments give much better results, at ILC at $1$ TeV with $2.5 ab^{-1}$ the predicted accuracy is $13\%$ \cite{Asner:2013psa}, and similarily at a $100$ TeV pp collider with $30 ab^{-1}$ of data \cite{Barr:2014sga}.
This is also the predicted accuracy we will use while discussing allowed parameter space in Section~\ref{sec:sphlaboundsmod}.

The situation is much simpler in the colored scalar model where both gluon fusion production and partial decay widths of the Higgs boson are modified due to loops including the new scalar.
We will express the relevant branching ratios as
\begin{equation}
\Gamma (h\rightarrow X)= \frac{\left|\mathcal{A}_X\right|^2}{\left|\mathcal{A}_X^{\rm SM} \right|^2} \Gamma (h\rightarrow X)^{\rm SM}.
\end{equation}
In what follows $N_c=3$ and the loop functions $F$, can be found in \cite{Djouadi:2005gj}. Charges and third components of isospin for SM fields are listed in Table~\ref{tab:eT3}.
The decay width to two gluons is given by,
\begin{equation}
\mathcal{A}_{g g}^{\rm SM} =
 \sum\limits_{i=d,s,b,u,c,t}F_{\frac{1}{2}}(\tau_i), \quad \quad
\mathcal{A}_{g g} = \mathcal{A}_{g g}^{\rm SM} + \frac{g_S}{4} \left(\frac{v_0}{m_S}\right)^2 F_{1}(\tau_S),
\end{equation} 
Similarly for the two photon decay width we have,
 \begin{eqnarray}
 \mathcal{A}_{\gamma \gamma}^{\rm SM} &=&
F_1(\tau_W)
+e_e^2\sum\limits_{i=e,\mu,\tau}F_{\frac{1}{2}}(\tau_i)
+N_c e_d^2 \sum\limits_{i=d,s,b}F_{\frac{1}{2}}(\tau_i)
+N_c e_u^2 \sum\limits_{i=u,c,t}F_{\frac{1}{2}}(\tau_i) \nonumber
\\
\mathcal{A}_{\gamma \gamma} &=&
\mathcal{A}_{\gamma \gamma}^{\rm SM} +
 N_c e_d^2  \frac{g_S}{4} \left(\frac{v_0}{m_S}\right)^2 F_{1}(\tau_S) .
\end{eqnarray} 
\begin{table}[t]
\centering
\begin{tabular}{c|c c c c c}
$f$  & $u$ & $d$ & $e$&
\\ \hline 
$e_f$ &  $\frac{2}{3}$& $-\frac{1}{3}$& $-1$&
 \\
$T_3^f$ & $\frac{1}{2}$& $-\frac{1}{2}$& $-\frac{1}{2}$& 
\end{tabular}
\caption{Charges and effective third isospin components.
}\label{tab:eT3}
\end{table}
Remaining decay widths are either very small or exist at tree level in the SM and thus their modification comes only from a small  loop correction shift.
The branching ratios are given by
\begin{equation}
B(h \rightarrow X)=\frac{\Gamma_X}{\sum\limits_i \Gamma_i}
\end{equation}
with the sum running over all decay channels.
We use the SM branching ratios given in \cite{Almeida:2013jfa}.
We can approximate the resulting prediction for signal strength modification by including only gluon fusion production mode, which at leading order gives
\begin{eqnarray}
\Delta \mu_X &=& \frac{\sigma B(h \rightarrow X)-\sigma^{\rm SM} B^{\rm SM}(h \rightarrow X)}{ \sigma^{\rm SM} B^{\rm SM}(h \rightarrow X)}=
 \frac{\sigma B(h \rightarrow X)}{ \sigma^{\rm SM} B^{\rm SM}(h \rightarrow X)}-1
  \\ \nonumber & \approx &
\frac{\sigma( g g \rightarrow h)}{\sigma^{\rm SM}( g g \rightarrow h)} \frac{Br(h \rightarrow X)}{B^{\rm SM}(h \rightarrow X)}-1
\approx 
\frac{\Gamma(h \rightarrow g g )}{\Gamma^{\rm SM}(h \rightarrow g g)} \frac{B(h \rightarrow X)}{B^{\rm SM}(h \rightarrow X)}-1.
\end{eqnarray}


The resulting modifications of the signal strength is dominated by the increased $gg\rightarrow H$  production cross-section. 
When comparing our prediction to the experimental sensitivities the most important limit comes from the $H \rightarrow WW$ signature. High statistics and good sensitivity at the LHC make this channel more important than $H\rightarrow \gamma \gamma$ which is less useful due to cancellation between increased production and decreased branching ratio. 
Still both of these modifications are large in the part of the parameter space predicting EWBG as we will discuss in the next section.
\section{Details of the phase transition} 
As the temperature of the universe drops below the critical temperature the minimum in which electroweak symmetry is broken becomes the global minimum of the potential. At this time the field is still in a homogeneous state in the symmetric local minimum, and  separated from the emerging global minimum by a potential barrier which is generated due to thermal fluctuations.
As the temperature drops and the barrier between vacua shrinks, bubbles of the broken symmetry vacuum begin to nucleate within the symmetric background due to thermal tunnelling. 
We will now shortly review the computational details of the phase transition.
The transition proceeds due to a thermal tunneling effect described by spontaneous nucleation of bubbles of the broken phase in the background consisting of a homogeneous configuration of the field still in the symmetric minimum. 
The bubble  nucleating due to a temperature fluctuation is a static $O(3)$ symmetric field configuration with action given by
\begin{equation} \label{eq:actionfunc}
S_3=4\pi \int dr  r^2\left[\frac{1}{2}\left(\frac{d\phi}{dr}\right)^2+V(\phi,T)\right].
\end{equation}
The probability of nucleation of a bubble per volume $\mathcal{V}$ is given by \cite{Linde:1981zj, Linde:1980tt}
\begin{equation}\label{eq:decaywidth}
\Gamma/\mathcal{V} \approx T^4 \exp\left(-\frac{S_3(T)}{T}\right).
\end{equation}
The corresponding equation of motion for the field takes the form
\begin{equation}\label{eqn:scalarEOM}
\frac{d^2 \phi}{dr^2}+\frac{2}{r}\frac{d \phi}{dr}+\frac{\partial V(\phi,T)}{\partial\phi}=0,
\end{equation}
with boundary conditions
\begin{equation}
\phi (r\rightarrow\infty)= 0 \mbox{\ \ and\ \ } \frac{d\phi(r=0)}{dr}=0. 
\end{equation}
We numerically solve the equation of motion \eqref{eqn:scalarEOM} using the full effective potential \eqref{eqn:Veff} and use an overshoot/undershoot algorithm to satisfy the boundary conditions.
This allows us to accurately compute the action \eqref{eq:actionfunc} and the decay width \eqref{eq:decaywidth}. The temperature of the phase transition $T_*$ depends also on the cosmological history, as we assume that the transition proceeds when at least one bubble appears in every horizon.
However, as we have shown in \cite{Lewicki:2016efe} this dependence is very weak and we will not describe it here in more detail.

\section{Modification of the cosmological history}

We will discuss a modification of cosmology on a very generic model which can effectively describe most of available cosmological models. 
We simply assume that the energy density of the universe has a new constituent $\rho_S$ that redshifts faster than radiation. 
The Friedmann equation, including the new component and radiation reads
\begin{equation}\label{eq:friedmann}
H^2=\left( \frac{\dot{a}}{a} \right)^2 = \frac{8\pi}{3 M_{p}^2}\left( \frac{\rho_N}{a^n} + \frac{\rho_R}{a^4}\right),
\end{equation}
with $n > 4$ for the new constituent.

The crucial point here is that there is an important experimental constraint coming from Big Bang Nucleosynthesis (see e.g. \cite{Cooke:2014,Agashe:2014kda}) that one can put on all models of this kind. To recreate observed abundances of light elements neutrons have to freeze-out cosmologically saving a precisely known fraction from decay, this gives us the Hubble rate at the temperature of roughly $1$ MeV when this process occurs. The observed rate is consistent with universe filled by SM radiation. However there is still some room within experimental uncertainty for an additional component which we will identify with $\rho_N$.
To obtain bounds on this additional contribution we translate the bound on the effective number of neutrinos directly to the modification of the Hubble rate \cite{Simha:2008zj},
\begin{equation}
\left. \frac{H}{H_R} \right|_{\rm BBN}=\sqrt{1+\frac{7}{43}\Delta N_{\nu_{\rm eff}}}.
\end{equation}
Where $\Delta N_{\nu_{\rm eff}}$ is the difference between the SM radiation $N=3.046$ and the observed  central value $N_{\nu_{\rm eff}}=3.28\pm 0.28$ \cite{Cooke:2014,Agashe:2014kda}.

The next step is to calculate the allowed modification of the Hubble rate at the temperature of the phase transition.
Since the new energy constituent does not interact with SM degrees of freedom, the usual 
relationship between temperature and scale factor holds, namely
\begin{equation}\label{eq:rhorad}
\frac{\rho_{R}}{a^4}=\frac{\pi^2}{30}g_* T^4,
\end{equation} 
where $g_*=106.75$ is the number of degrees of freedom contributing to the plasma and it can be approximated as a constant in the SM around the temperature of the phase transition.

Moving towards earlier times the new component contribution quickly comes to dominate the total energy density in \eqref{eq:friedmann}. Thus a simplified form of the Friedman equation neglecting $\rho_R$ can be used which leads to a very simple result,
\begin{equation}
\frac{H}{H_R} =
\sqrt{\left(\left. \frac{H}{H_R} \right|_{\rm BBN} \right)^2-1}
\left(\left(\frac{g_{*}}{g_{\rm BBN}}\right)^{\frac{1}{4}}\frac{T_*}{T_{\rm BBN}}\right)^{\frac{n-4}{2}},
\end{equation} 
where all values without the subscript ``BBN" are calculated at the phase transition temperature $T_*$, while $T_ {\rm BBN}=1$\mev and $g_{\rm BBN}=10.75$. 

The resulting maximal modification of the expansion rate in the relevant temperature range $T\in [100,150]\gev$ is shown in Figure~\ref{HHRmaxandvoverT}. The value $n=6$ is the boundary for a simple interpretation of the new component as a perfect fluid. However, even conservatively neglecting more uncommon scenarios with $n>6$ we can get an expansion more than 5 orders of magnitude faster than in the standard case without violating any observational bounds.

\section{Evolution of primordial inhomogeneities}

The primordial inhomogeneities of matter fields and the metric tensor are the seeds of the large scale structure of the Universe. We observe them as anisotropies of the cosmic microwave background radiation \cite{Ade:2015xua}, from which one concludes that inhomogeneities are very small (the observed deviation from the average CMB temperature is of order $\Delta T/T \sim 10^{-5}$) and therefore linear. Let us consider the evolution of gauge-invariant scalar metric perturbations. Assuming the equation of state $p = p(\rho,S)$, where $p$, $\rho$ and $S$ are pressure, energy and entropy densities respectively, one finds $\delta p = c_s^2 \delta \rho + \tau \delta S $, where $\delta p$, $\delta \rho$ and $\delta S$ are gauge invariant perturbations 
, $c_s$ is the speed of sound and $\tau = (\partial p/\partial S)_\rho$. Assuming the lack of anisotropic pressure one finds 
\begin{eqnarray}
\Phi'' + 3(1+c_s^2)\mathcal{H}\Phi' - c_s^2 \Delta \Phi + (2\mathcal{H}'+(1+3c_s^2)\mathcal{H}^2)\Phi &=& \frac{a^2}{2}\tau \delta S \, , \label{eq:EOMpert} \\
 2\Delta\Phi - 6\mathcal{H}(\Phi'+\mathcal{H} \Phi) &=& a^2 \delta\rho \, ,
\end{eqnarray}
where $' = \frac{d}{d\eta}$, $a(\eta)d\eta = dt$ is the conformal time and $\mathcal{H} = a'/a$ is the conformal Hubble parameter and $\Phi$ is a gauge-invariant scalar metric perturbation. It is convenient to analyze the Fourier modes of inhomogeneities, which at the level of equations of motion gives $\Delta \Phi \to -k^2 \Phi$. 

The evolution of inhomogeneities can be analyzed in two limits: for super-horizon scales, when $k\eta \ll 1$, or scales deep under the Hubble horizon, when $k\eta \gg 1$. Let us assume that the universe is filled with radiation and the additional component scales as $\rho \propto a^{-6}$. For adiabatic perturbations ($\delta S = 0$) and $k \eta \ll 1$ one finds
\begin{eqnarray}
a(\eta) = \sqrt{\frac{\eta}{\eta_\star}\left(2+\frac{\eta}{\eta_\star}\right)} \, , \label{eq:aofeta} \\
\Phi(\eta) = \frac{1}{12}\left(8 A-\frac{3 B}{\left(\eta \left/\eta _\star\right.\right)^2}+\frac{4 A+3 B}{\left(2+\eta \left/\eta _\star\right.\right)^2}\right) \, ,
\end{eqnarray}
where $A$ is a constant set by the normalization of inhomogeneities, $B$ is the decaying mode, $\eta_\star : =\eta_{eq}/(\sqrt{2}-1) = (\rho_{eq}/24)^{-1/2}$, and $\eta_{eq}$ is the conformal time at which radiation and the additional component have the same energy density equal to $\rho_{eq}$.

For the modes below the horizon one cannot find an analytical solution for the $n=6$ plus radiation system. However, it is easy to show that for the domination of an additional component with $n \neq 3$ in the $k \eta \gg 1$ limit one finds $\Phi$ and $\delta$ oscillating with the period $\sim 1/(k\sqrt{w})$ and an amplitude that evolves like  
\begin{equation}
\Phi \propto \eta^{-\frac{n}{n-2}} \propto a^{-\frac{n}{2}} \, , \qquad \delta \propto \eta^{\frac{n-4}{n-2}} \propto a^{\frac{n-4}{2}} \, ,
\end{equation}
where $\delta := \delta \rho/\rho$. This means that for any $n > 4$ the amplitude of $\delta$ increases under the horizon. In particular, for $n=6$ the $\delta$ grows as $\sqrt{\eta} \sim a(\eta)$, which is similar to the result due to dust domination, when $\delta \propto a$ for $k\eta \gg 1$. In the dust domination case this leads to growth of large scale structure in the universe, emerging from primordial inhomogeneities. However, this occurs at the era of last scattering ($T\leq eV$) during which scales of order of the Hubble horizon correspond to the size of galaxies. 
The additional component dominates much earlier ($T \gg MeV$) and the modes that satisfy $k \eta \gg 1$ correspond to scales much smaller than any cosmologically significant distance. 
Such modes will be strongly suppressed during the era of last scattering due to diffusion damping also known as Silk damping \cite{Silk:1967kq}. This effect is based on diffusion of primordial inhomogeneities by photons during the recombination era. As a result the inhomogeneities related to small scales are being exponentially washed out rendering our model safe and inert with respect to astrophysical experimental tests.


\section{Cosmological modification of the sphaleron bound}\label{sec:sphlaboundsmod}

The most important condition necessary for electroweak baryogenesis is decoupling of the sphaleron processes after the phase transition has taken place. If this bound is not fulfilled all the asymmetry created during the phase transition will be subsequently washed away.
The sphaleron processes in the SM are $SU(2)$ gauge interactions and are heavily suppressed once this symmetry is broken. 
A simple  criterion for this decoupling requires 
the sphaleron rate to be smaller than the Hubble rate 
\begin{equation}
\Gamma_{\rm Sph} = T^4 \mathcal{B}_0 \frac{g}{4 \pi} \left(\frac{v}{T}\right)^7 \exp \left( -\frac{4 \pi}{g}\frac{v}{T} \right)\lessapprox H,
\end{equation}
where $v$ is the vev of the Higgs field at the time of the phase transition and the constant $\mathcal{B}_0$ depends on SM couplings and contains loop corrections to the sphaleron rate.
Calculation of $\mathcal{B}_0$ is difficult and different values are used in literature leading to slightly different bounds on $v/T$ \cite{Katz:2014bha,Quiros:1999jp,Funakubo:2009eg,Fuyuto:2014yia}, here we simply use a value that leads to the standard bound $v/T \leq 1$ for the Hubble rate predicted by SM radiation $H=H_R$.

\begin{figure}[t]
\begin{center}
\includegraphics[height=6.5cm]{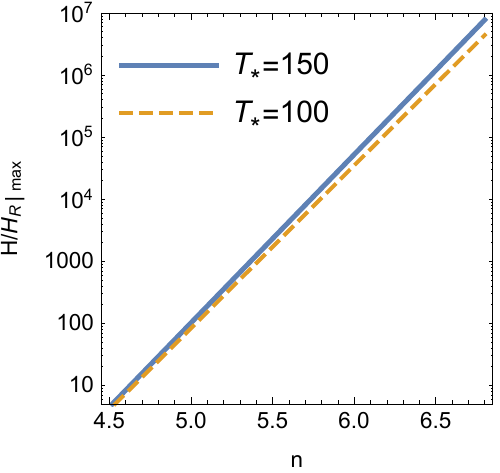} 
\includegraphics[height=6.5cm]{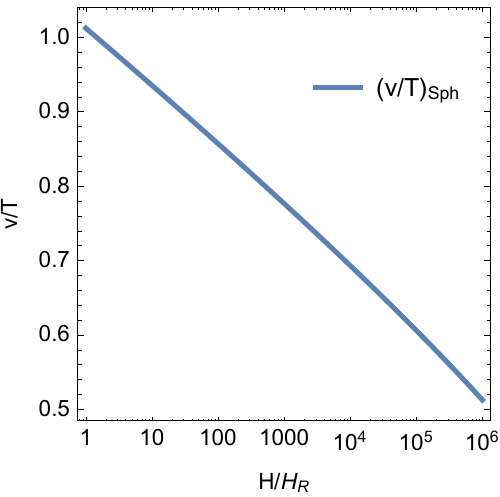} 
\end{center}
\caption{
Left panel: maximal modification of the Hubble parameter calculated at the nucleation temperatures $T_*=100$ GeV and $T_*=150$ GeV, as a function of the parameter
 $n$ which determines our cosmological model.
 Right panel: the modified required value of the ratio $v/T$ (evaluated at $T_*$) needed to preserve the baryon asymmetry created during the transition. This modified value from cosmological freeze-out of the sphaleron processes and is a function of the expansion rate normalized to the standard case of radiation dominated expansion.  
\label{HHRmaxandvoverT}
}
\end{figure}

This brings us to the main point of this paper, the dependence of sphaleron decoupling on cosmological history.
We already discussed how the expansion rate in the early universe can be increased in the early universe and now it is straightforward to see how our required $(v/T)_{\rm Sph}$ decreases with the faster expansion. The result is shown in Figure~\ref{HHRmaxandvoverT}.
We are now ready to combine this data with detailed information on the phase transition in our models to obtain the allowed parameter space and accelerator constraints.

\begin{figure}[t] 
\begin{center}
\includegraphics[height=7.5cm]{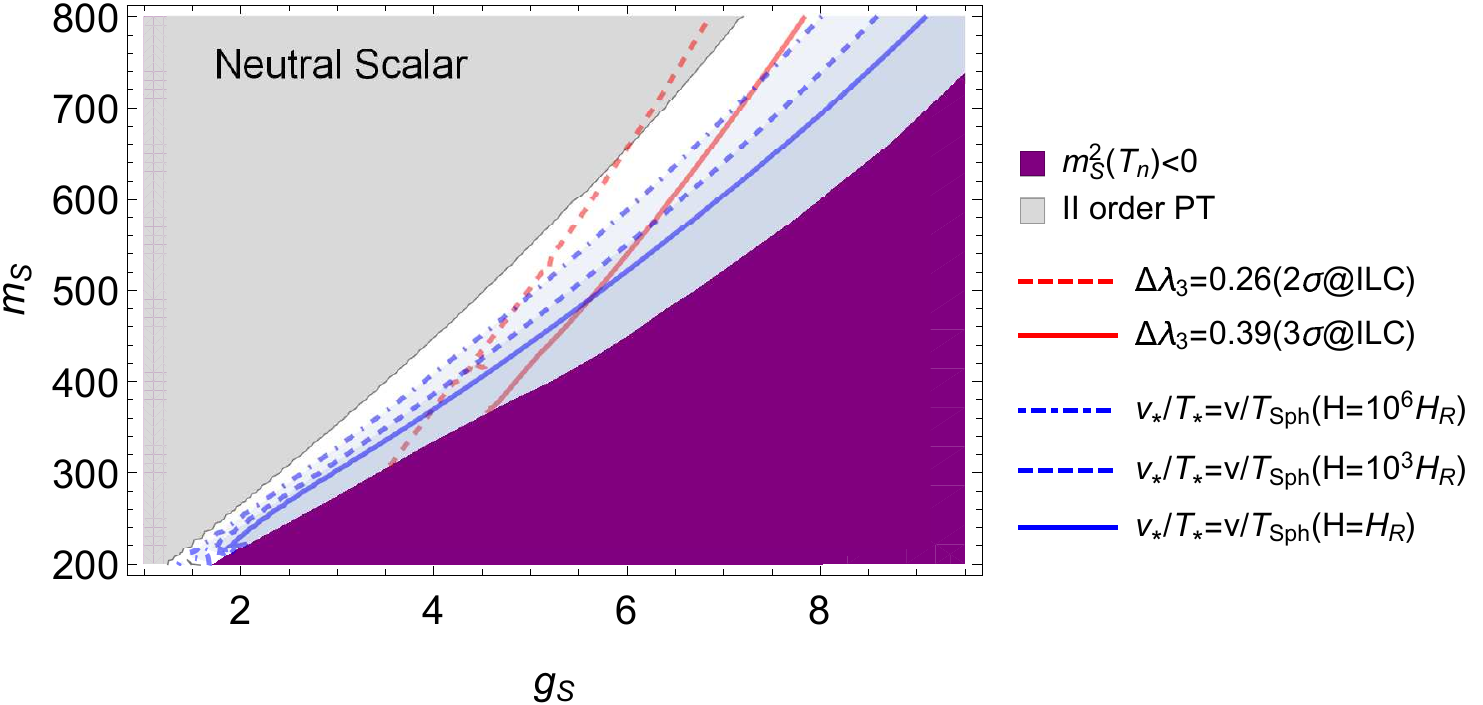} 
\end{center}
\caption{
Region in parameter space of the neutral scalar mass $m_S$ and the Higgs mixing $g_S$ predicting successful baryogenesis together with predicted ILC experimental constraints on the triple Higgs coupling $\lambda_3$. Three different allowed (blue) regions correspond to standard cosmological history and expansion during the phase transition accelerated $10^3$ times and $10^6$ times.
\label{fig:NSExcll3}
}
\end{figure}

\begin{figure}[t] 
\begin{center}
\includegraphics[height=7.5cm]{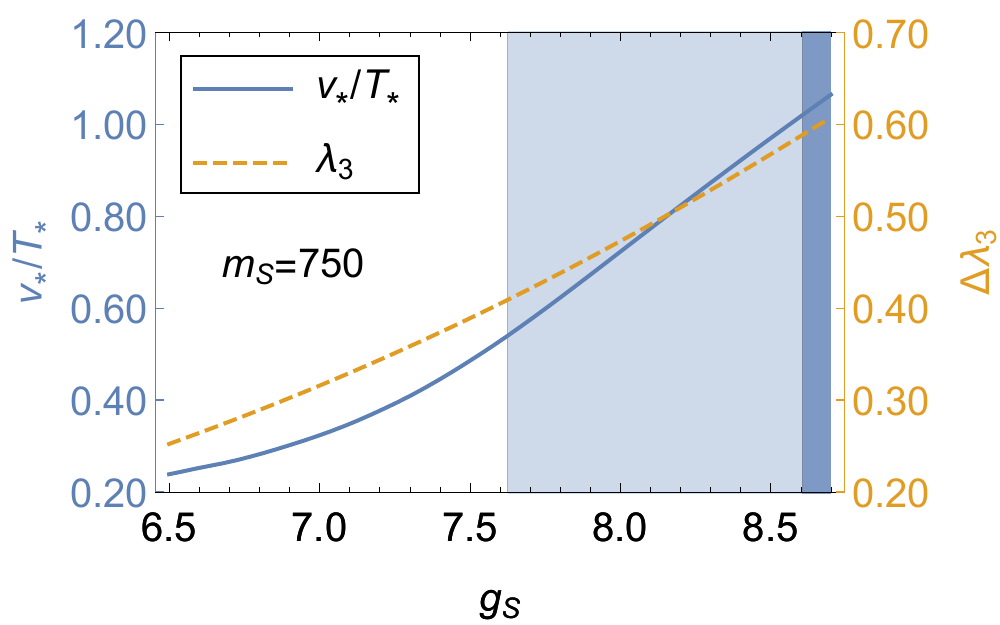} 
\end{center}
\caption{
Values of neutral scalar mixing with the Higgs $g_S$ allowing successful baryogenesis due to modified cosmological history (light blue region) together with modification of the triple Higgs coupling $\lambda_3$. The dark blue region, and the region to the right with even larger mixing, predicts a strong enough phase transition without cosmological modification.
\label{fig:NSvTallowed}
}
\end{figure}

Figure~\ref{fig:NSExcll3} shows the parameter space of the neutral scalar model  in coupling $g_S$ and mass $m_S$, highlighting the parts of parameter space allowing electroweak baryogenesis and highlighting regions made plausible due to the faster expansion rate of the universe during the phase transition.  
Even though the excitement about a possible signal at $750$ GeV has passed, this part of the parameter space bears a lot of significance in our model as above $m_S=700\gev$ EWBG will always result in an at least $3\sigma$ deviation in $\lambda_3$ observed at ILC.
Figure~\ref{fig:NSvTallowed} shows the relevant values of the coupling for $m_S=750\gev$ and highlights the light blue region which is acceptable due to the cosmological modification.
The important conclusion here is that ILC will be able to exclude the very simple model including only a new neutral scalar for scalar masses above $\approx 700\gev$ if we require successful EWBG.

\begin{figure}[t] 
\begin{center}
\includegraphics[height=7.5cm]{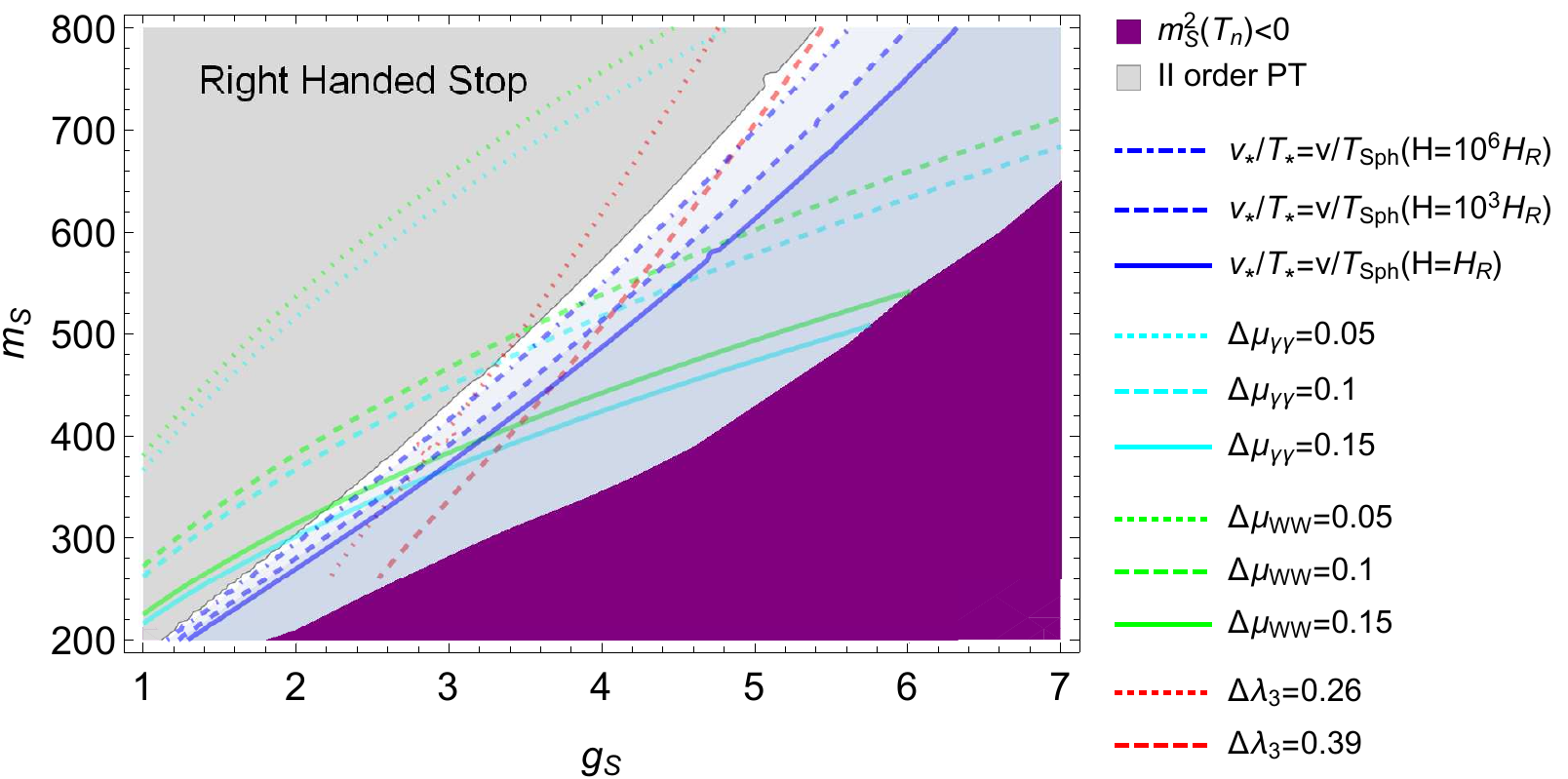} 
\end{center}
\caption{
Region in parameter space of new scalars mass $m_S$ and the Higgs mixing $g_S$ predicting successful baryogenesis together with modification of Higgs boson signal strength in $\gamma \gamma$ and $W W$ channels and modification of the triple Higgs coupling $\lambda_3$. Three different allowed (blue) regions correspond to standard cosmological history and expansion during the phase transition accelerated $10^3$ times and $10^6$ times.
\label{fig:RHSExclggWW}
}
\end{figure}

\begin{figure}[t] 
\begin{center}
\includegraphics[height=7.5cm]{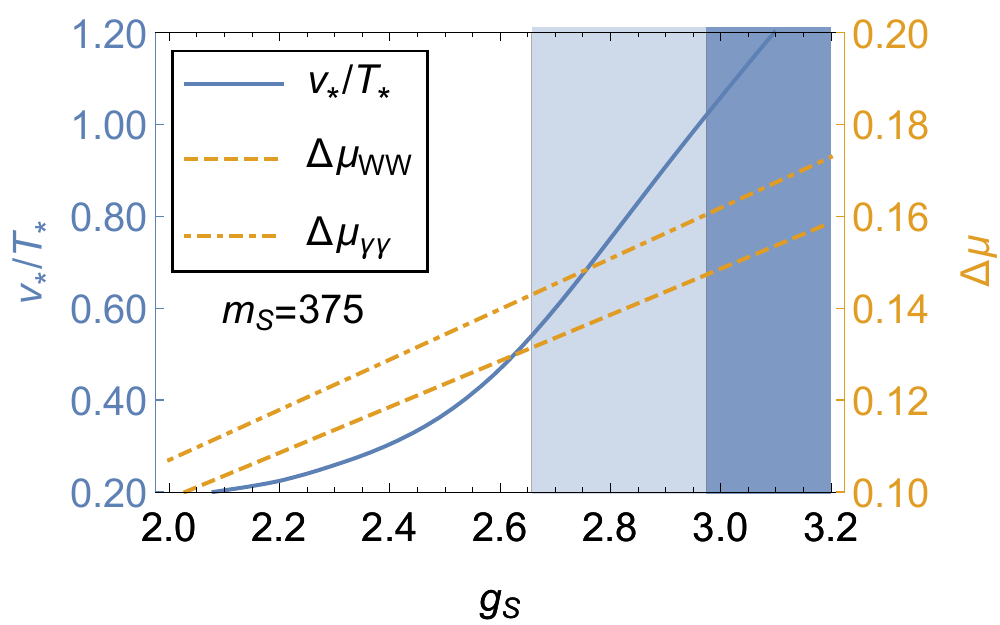} 
\end{center}
\caption{
Values of the Higgs mixing $g_S$ allowing successful baryogenesis due to modified cosmological history (light blue region) together with modification in Higgs boson signals in the right handed stop model.  The dark blue region (and even larger mixing region) predicts a strong enough phase transition without cosmological modification.
\label{fig:RHSvTallowed}
}
\end{figure}

Figure~\ref{fig:RHSExclggWW} shows the allowed region in the parameter space of the right handed stop model. Due to its color degrees of freedom this model requires smaller couplings than the previous one in order to facilitate EWBG. However, here new particles running in loops modify both Higgs production and decay, making it easier to probe using Higgs precision data.
We can see that even after HL-LHC we should have more than a $2\sigma$ deviation if the mass of the scalar is less than $375\gev$. This particular mass value is also interesting as a very similar model with this mass was suggested as a possible source of the $750\gev$ signal \cite{Kats:2016kuz}.
Figure~\ref{fig:RHSvTallowed} shows the relevant values of the coupling for $m_S=375\gev$, highlighting the light blue region acceptable due to the cosmological modification.
 
The key observation here is that the Higgs signal strength measurements are most useful at low masses and thus are complementary in excluding this model with $\lambda_3$ modification which is most useful at high masses. In fact after the run of ILC at $1$ TeV with $2.5 ab^{-1}$ data the predicted accuracy in $\mu_{WW}$ is $1.6\%$ \cite{Asner:2013psa} which together with measurement of $\lambda_3$ would either find evidence for a RHS interpretation of EWBG or exclude the entire parameter space of the RHS model that realizes EWBG.

\section{Gravitational waves detection}

Gravitational waves are produced during a first-order phase transition by three main mechanisms. These are collisions of bubble walls \cite{Kamionkowski:1993fg,Huber:2008hg}, sound waves \cite{Hindmarsh:2013xza,Hindmarsh:2015qta} and magnetohydrodynamical turbulence \cite{Caprini:2009yp} in the plasma after the collisions.
All of these contributions can be calculated knowing details of the phase transition. Two parameters describing the transition are particularly useful.
The first one is the ratio of latent heat released after the phase transition to the total energy density, which in the usual case is just radiation density,
\begin{equation}
\alpha= \left.
\frac{1}{\rho_{\rm tot}}\left(T\frac{d V_{eff}^{\rm min}}{d T}-V_{eff}^{\rm min}\right)
\right|_{T=T_*} ,
\end{equation}
where we already used the fact that the value of the field and the potential in the symmetric minimum is equal to zero, and the superscript min refers to the symmetry breaking minimum. 
The second one is the inverse time of the phase transition defined as
\begin{equation}
\frac{\beta}{H}=T_* \left. \left( \frac{d}{dT} \frac{S_3(T)}{T} \right)\right|_{T=T_*}.
\end{equation}

For bubble collision contribution the peak frequency reads~\cite{Huber:2008hg}
\begin{equation}
f_{col}=
16.5\times 10^{-6} \frac{0.62}{v_b^2-0.1 v_b+1.8}\frac{\beta}{H}
\frac{T_*}{100} \left(\frac{g_*}{100}\right)^{\frac{1}{6}} {\rm Hz}
\end{equation}
while the total energy reads
\begin{equation}
\Omega h^2_{col}=1.67\times 10^{-5}\left(\frac{\beta}{H}\right)^{-2}
\frac{0.11 v_b^3}{0.42+v_b^2}
\left(\frac{\kappa \alpha }{1+\alpha }\right)^2 
\left(\frac{g_*}{100}\right)^{-\frac{1}{3}}
\frac{3.8 \left(f/f_{col}\right)^{2.8}}{1+2.8 \left(f/f_{col}\right)^{3.8}}
\end{equation}
where the bubble wall velocity and efficiency factor $\kappa$ are both functions of $\alpha$
\begin{equation}
v_b=\frac{1/ \sqrt{3}+\sqrt{\alpha ^2+2 \alpha /3}}{1+\alpha }, \quad
\kappa=\frac{1}{1+0.715 \alpha }\left(0.715 \alpha +\frac{4}{27} \sqrt{\frac{3 \alpha }{2}}\right).
\end{equation}
For points with large $\alpha$ the energy deposited into fluid saturates at \cite{Espinosa:2010hh}
\begin{equation}
\alpha_{\infty}=4.9\times 10^{-3} \left(\frac{v_*}{T_*}\right)^2
\end{equation}
and the fraction of energy going to fluid motion reads
\begin{equation}
\kappa=\frac{\alpha_{\infty}}{\alpha}\left( \frac{\alpha_{\infty}}{0.73+0.083\sqrt{\alpha_{\infty}}+\alpha_{\infty}} \right)
\end{equation}
corresponding to the so called runaway configurations \cite{Bodeker:2009qy,Megevand:2009gh}.
This effect diminishes the signal in the majority of parameter space allowed by EWBG.

Motion of the fluid after bubble collisions results in a sound wave contribution to the gravitational waves. The peak frequency is given by \cite{Hindmarsh:2013xza,Hindmarsh:2015qta}
\begin{equation}
f_{sw}=1.9 \times 10^{-5} \frac{\beta}{H} v_b^{-1} \frac{T_*}{100}\left({\frac{g_*}{100}}\right)^{\frac{1}{6}} {\rm Hz }
\end{equation}
while the total energy reads
\begin{equation}
\Omega h^2_{sw}=2.65 \times 10^{-6}\left(\frac{\beta}{H}\right)^{-1}
\left(\frac{\kappa \alpha }{1+\alpha }\right)^2 
\left(\frac{g_*}{100}\right)^{-\frac{1}{3}}
v_b
\left(f/f_{sw}\right)^3 \left(\frac{7}{4+3 \left(f/f_{sw}\right)^2}\right)^{7/2}
\end{equation}
The last possible contribution to gravitational wave signals is MHD turbulence in the ionized plasma. This signal is peaked at~\cite{Caprini:2009yp}
\begin{equation}f_{turb}=2.7  \times 10^{-5}
\frac{\beta}{H} v_b^{-1} \frac{T_*}{100}\left({\frac{g_*}{100}}\right)^{\frac{1}{6}} {\rm Hz }
\end{equation} 
and its contribution to the gravitational wave spectrum reads
\begin{equation}
\Omega h^2_{sw}=3.35 \times 10^{-4}\left(\frac{\beta}{H}\right)^{-1}
\left(\frac{\epsilon \kappa \alpha }{1+\alpha }\right)^{\frac{3}{2}} 
\left(\frac{g_*}{100}\right)^{-\frac{1}{3}}
v_b
\frac{\left(f/f_{sw}\right)^3\left(1+f/f_{sw}\right)^{-\frac{11}{3}}}{\left(1+8\pi f a_0/(a_* H_*)\right)},
\end{equation}
where $\epsilon\approx 0.05$.
Contrary to the previous contributions (bubble collisions and sound waves) here we have an explicit dependence on the Hubble rate. In our model this contribution is always a few orders of magnitude smaller than the previous two.
\begin{figure}[t] 
\begin{center}
\includegraphics[height=5.7cm]{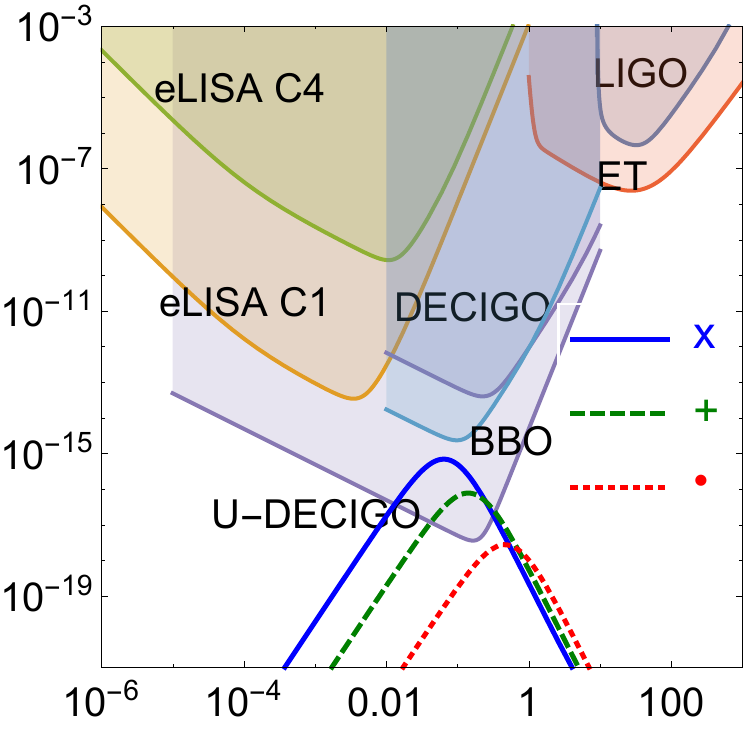} 
\includegraphics[height=5.7cm]{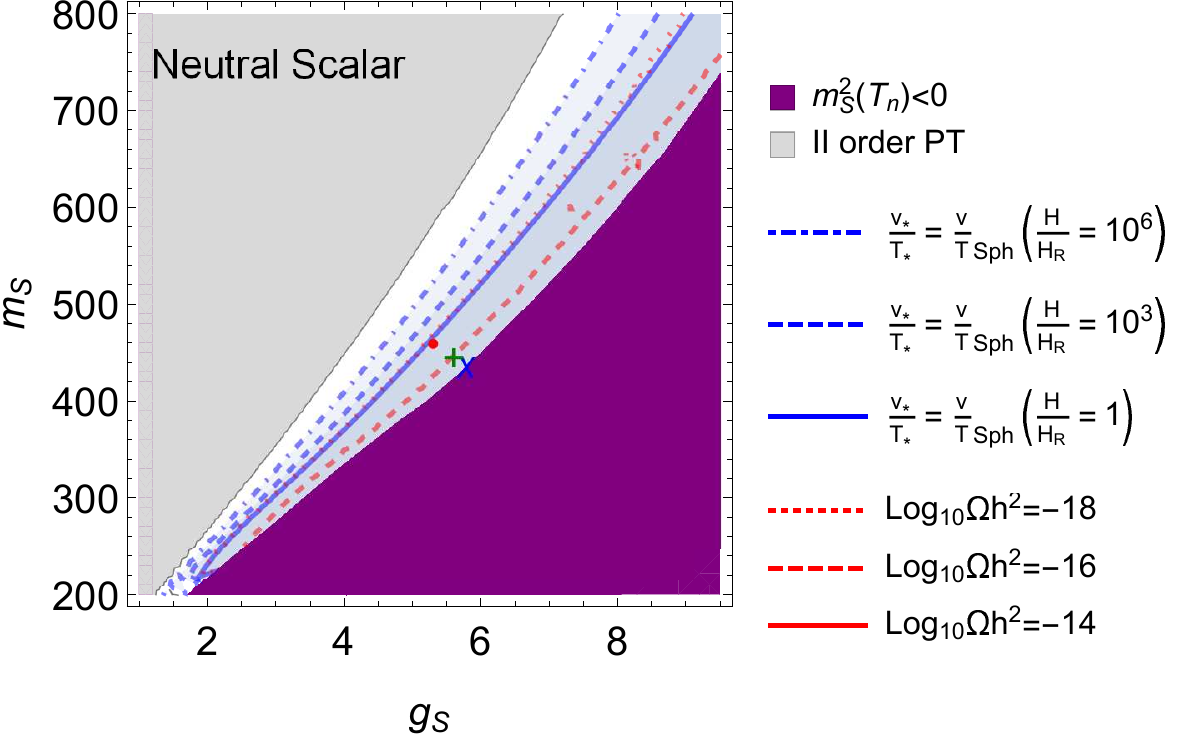} 
\end{center}
\caption{
Right panel: parts of parameter space allowing successful EWBG (as in Figure~\ref{fig:RHSExclggWW}) together with contours showing where emitted GW signal could be detected at BBO, DECIGO and eLISA.
Left panel: examples of GW signals corresponding to points in the parameter space marked on the right hand side plot.
\label{fig:NSGWplot}
}
\end{figure}
\begin{figure}[t] 
\begin{center}
\includegraphics[height=5.7cm]{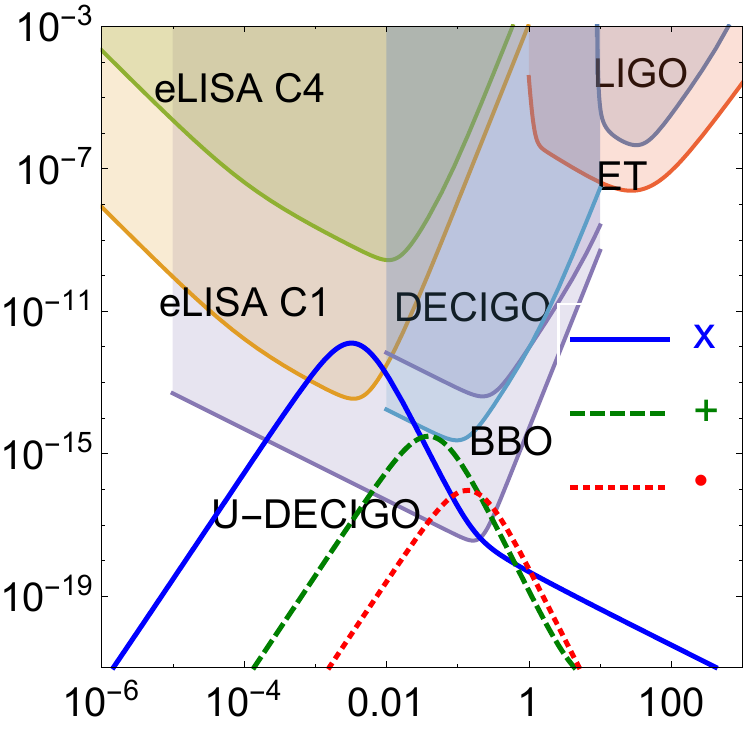} 
\includegraphics[height=5.7cm]{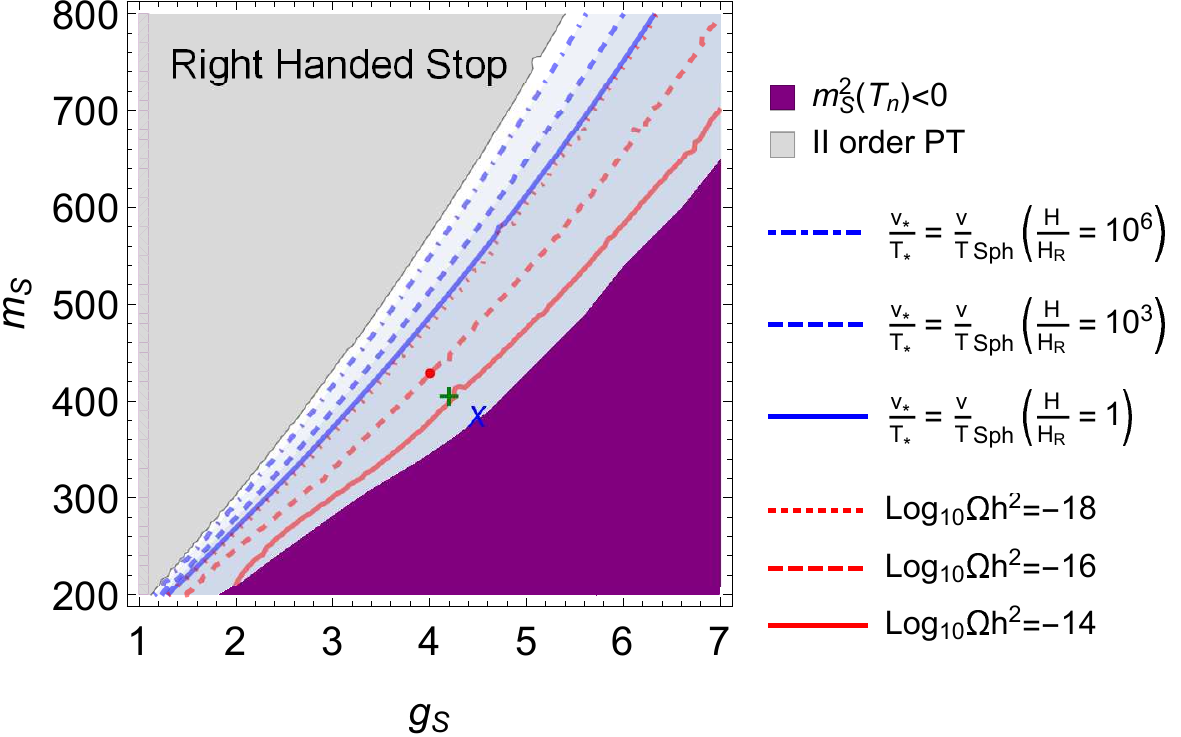} 
\end{center}
\caption{
Right panel: parts of parameter space allowing successful EWBG (as in Figure~\ref{fig:RHSExclggWW}) together with strength of emitted GW signal.
Left panel: examples of GW signals corresponding to points in the parameter space marked on the right hand side plot. 
\label{fig:RHSGWplot}
}
\end{figure}

Gravitational wave signal grows with the strength of the phase transition which is shown in right panel of Figures~\ref{fig:NSGWplot} and~\ref{fig:RHSGWplot}. Future experiments aLIGO \cite{TheLIGOScientific:2014jea}, ET \cite{Hild:2010id}, eLISA (using most and least promising configurations C1 and C4) \cite{Klein:2015hvg,Caprini:2015zlo}, DECIGO, BBO \cite{Yagi:2011wg} and Ultimate-DECIGO \cite{Kudoh:2005as} predict the possibility of detection only in the region of strong phase transition in which standard radiation dominated cosmology suffices for baryogenesis.

Cosmological modification can enable baryogenesis in a much larger region where the phase transition is weaker.
Gravitational waves redshift just as radiation and their value today is set by the ratio of temperatures today and during the transition, which as we discussed is modified very weakly.
As for the production, usual approximations assume that the expansion of the universe \cite{Grojean:2006bp} and all constant homogeneous backgrounds can be neglected \cite{Kamionkowski:1993fg}, and therefore the result should not be modified in this leading approximation.
The only modification comes from the modified relation between time and temperature which makes the phase transition much shorter and as a result less sources contribute to the signal at any given point.
In the simplest approximation neglecting all scales arising due to the phase transition itself this leads to a scaling of amplitude as $\Omega_{GW} \propto (H/H_R)^{-2}$ \cite{Chung:2010cb}. However, a more detailed analysis including such scales would probably lead to results between this simple scaling and the usual radiation dominated case.
 The peak frequency of produced gravitational waves changes as $f\propto (H/H_R)$ and our cosmological modification would push the signals toward higher frequencies making their detection even more difficult.
However, the modification is needed only in the region where the phase transition is weak and the energy carried by gravitational waves too small for detection anyway.
Thus in the region of parameter space where EWBG is enabled by modified expansion the gravitational wave signal will also be very weak leaving little hope for detection. 

Thus we can conclude that observation of a germane deviation from the SM at the ILC without any corresponding gravitational wave signal can point to modified cosmologies if these signals are to also help explain baryogenesis. 

\section{Conclusions}

In this paper we studied detection possibilities for simple EWBG models that include only one new scalar with a possible modified cosmological history.
To this end we used a very generic model  to modify the cosmological history, which introduced a new energy density constituent which redshifted away before nucleosynthesis. 

We carefully computed the details of the EW phase transition going beyond the oft-used
critical temperature approximation. 
This allowed us to accurately compute the gravitational wave signal produced during the phase transition as the degeneracy of the minima of the potential during the transition plays a critical role there.

We also described the modification of $SU(2)$ sphalerons of the Standard Model due to the modified cosmological history. 
The main effect comes from cosmologically modified freezout of the sphaleron processes after the phase transition.
This has a severe impact on the corresponding detection range for collider experiments changing the exclusion range by as much as a few hundred GeV.

Next we computed the gravitational wave signals produced during the phase transition in our model. These turn out important only in the region where the phase transition is strong enough to allow baryogenesis without a cosmological modification.
Thus we conclude that observation of a modification of the Higgs observables in future collider experiments without a corresponding gravity wave signal could point to scenarios with a modified cosmological history. 

\section*{Acknowledgements}
We would like to thank T. Konstandin for useful discussions.
MA was supported by the Iuventus Plus grant No. 0290/IP3/2016/74 from the Polish Ministry of Science and Higher Education. 
ML is grateful to the DESY theory division for their hospitality and partial
support  during  the final stages of  this  work.
ML was supported in part by the Polish National Science Centre under research grant 2014/13/N/ST2/02712 and doctoral scholarship number 2015/16/T/ST2/00527 and under MNiSW grant IP2015 043174.
JDW was supported in part by the U.S.\ Department of Energy (DE-SC0007859) and the Alexander von Humboldt Foundation (Humboldt Forschungspreis). 
\bibliographystyle{JHEP}
\bibliography{EWPTandCOSMObibliography}

\end{document}